# On the Module of Internet Banking System

Hamdan.O.Alanazi, Rami Alnaqeib, Ali K.Hmood, M.A.Zaidan , Yahya Al-Nabhani

**Abstract** – Because of the speed, flexibility, and efficiency that it offers, the Internet has become the means for conducting growing numbers of transactions between suppliers and large international corporations. In this way, the Internet has opened new markets to the world and has accelerated the diffusion of knowledge. The meaning of Internet markets or online business has been widely used in these days. The success of the business depends on its flexibility, availability and security. Since that the web-based systems should have a special way to design the system and implement it. Nowadays, the Internet Banking System widely used and the banks looking to provide the best quality system with highly available, fast response, secure and safe to use. The Unified Modelling Language (UML) is the uniquely language which is used to analyse and design any system. In this paper, the UML diagrams has been proposed to illustrate the design phase for any banking system. The authors, presented two types of architecture which is used for the Internet Banking System.

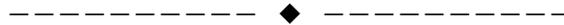

## 1. INTRODUCTION

In the recent years there has been explosion of Internet-based electronic banking applications (Liao & Cheung, 2003). Beckett, Hewer & Howcroft (2000) states that the emergence of new forms of technology has created highly competitive market conditions for bank providers. However, the changed market conditions demand for banks to better understanding of consumers' needs [1].

Liao et al. (2003) stress that the success in Internet banking will be achieved with tailored financial products and services that fulfill customer' wants, preferences and quality expectations. Mattila (2001) concedes that customer satisfaction is a key to success in Internet banking and banks will use different media to customize products and services to fit customers' specific needs in the future. Liao et al. (2003) suggest that consumer perceptions of transaction security, transaction accuracy, user friendliness, and network speed are the critical factors for success in Internet banking. From this perspective, Internet banking includes many challenges for human computer interaction (HCI) [2].

Hiltunen et al (2004) have remarked that there are at least two major HCI challenges in Internet banking. The first challenge is related to the problem how to increase the number of services of Internet banking and simultaneously guarantee the quality of service for individual customers [3]. The second challenge is related to the problem how to understand customer's needs, translate them into targeted content and present them in a personalized way in usable user interface. Hiltunen et al. (2004) imply that Internet banking research will concentrate more on HCI factors in the future [4]. Recently, Lindgaard & Dudek (2003) emphasize that now is an ideal time for HCI researchers to analyse user satisfaction, because there is growing interest in how to attract and increase the number of online customers in e-business and e-commerce. Lindgaard et al. (2003) stress that HCI researchers should reveal a structure of user satisfaction, determine how to evaluate it and conclude how it is related to the overall user experience of online customers. The concept of electronic banking has been defined in many ways (e.g. Daniel, 1999). According to Karjaluoto (2002) electronic banking is a construct that consists of several distribution channels. Daniel (1999) defines electronic banking as the delivery of banks' information and services by banks to customers via different delivery platforms that can be used with different terminal devices such as a personal computer and a mobile phone with browser or desktop software, telephone or digital television [5].

The system models are abstract view of a system that ignores some system details. Complementary system models can be developed other information about the system. And they are graphical representations that describe business processes, the problem to be solved and our system is to be developed.  This reading may be difficult to interpret. It is presented to cover the entire concept of process modeling using the specific language of systems design. In the homework for this module and in a separate module reading (the Demonstration Project), you will see these concepts applied to real situations which will help you integrate the concepts [6].

System models play an important role in systems development. Because system analysts are dealing with unstructured problems (and end-users, too, deal with them), there needs to be a systematic, logic-based approach to converting a real-world problem, no matter how vague, into a representation (or model) that captures the main points and relationships so that the problem can be analyzed. A model is a representation of the designer's interpretation of reality. Models can be building for existing systems as a way to understand better those systems, or for proposed systems as a way to document the organizational and information requirements or technical designs [7].

## 2. CONTEXTS DIAGRAM

A System Context Diagram is the highest level view of a system, similar to Block Diagram, showing a (normally software-based) system as a whole and its inputs and outputs from/to external factors. The Context Diagrams show the interactions between a system and other actors with which the system is designed to face. They are also typically drawn using labeled boxes to represent each of the external entities and another labeled box to represent the system being developed. The relationship is drawn as a line between the entities and the system being developed [7].

Context Diagram is a data flow diagram showing data flows between a generalized application within the domain and the other entities and abstractions with which it communicates. One thing that differentiates the use of data flow diagrams in domain analysis from other typical uses is that the variability of the data flows across the domain boundary must be accounted for with either a set of diagrams or text describing the differences [8].

Before we construct the actual process model, we need to establish initial project scope. A project is scope defines what aspect of the business a system or application is supposed to support. It also defines how the system or application being modeled must interact with other systems and the business as a whole. A project's scope is documented with a context diagram [8].

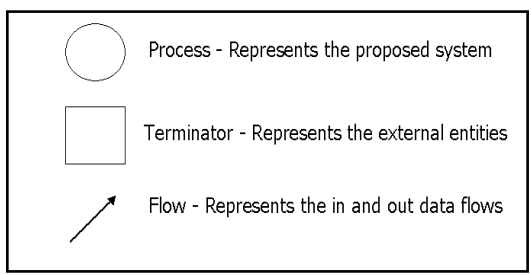

Fig 1. Notation for context Diagram

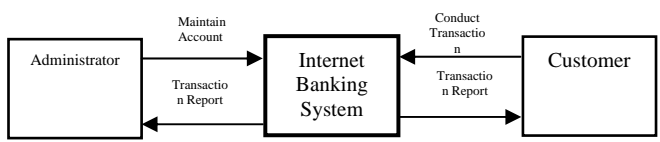

Fig 2. Context Diagram of internet banking system

## 3. SEQUENCE DIAGRAM

UML sequence diagrams model the flow of logic within your system in a visual manner, enabling you both to document and validate your logic, and are commonly used for both analysis and design purposes.

Sequence diagrams are the most popular UML artifact for dynamic modeling, which focuses on identifying the behavior within your system. Sequence diagrams, along with class diagrams and physical data models are in experts' opinion the most important design-level models for modern business application development.

The Message Sequence Chart technique has been incorporated into the Unified Modeling Language (UML) diagram under the name of Sequence Diagram. A sequence diagram shows, as parallel vertical lines, different processes or objects that live simultaneously, and, as horizontal arrows, the messages exchanged between them, in the order in which they occur. This allows the specification of simple runtime scenarios in a graphical manner. The dotted lines extending downwards indicate the timeline, time flows from top to bottom. The arrows represent messages (stimuli) from an actor or object to other objects [9].

The UML 2.0 Sequence Diagram supports similar notation to the UML 1.x Sequence Diagram with added support for modeling variations to the standard flow of events. If the lifeline is that of an object, it is underlined (if not it is a role). Note that leaving the instance name blank can represent anonymous and unnamed instances. In order to display interaction, messages are used. These are horizontal arrows with the message name written above them. Solid arrows with full heads are synchronous calls, solid arrows with stick heads are asynchronous calls and dashed arrows with stick heads are return messages. This definition is true as of UML 2, considerably different from UML 1.x. Activation boxes, or method-call boxes, are opaque rectangles drawn on top of lifelines to represent that processes are being performed in response to the message. Objects calling methods on themselves use messages and add new activation boxes on top of any others to indicate a further level of processing. When an object is destroyed (removed from memory), an X is drawn on top of the lifeline, and the dashed line ceases to be drawn below it. It should be the result of a message, either from the object itself, or another. A message sent from outside the diagram can be represented by a message originating from a filled-in circle. A UML diagram may perform a series of steps, called a super step, in response to only one external stimulus [10].

Sequence diagrams are typically used to model:

- **Usage scenarios**. A usage scenario is a description of a potential way your system is used. The logic of a

usage scenario may be part of a use case, perhaps an alternate course. It may also be one entire pass through a use case, such as the logic described by the basic course of action or a portion of the basic course of action, plus one or more alternate scenarios. The logic of a usage scenario may also be a pass through the logic contained in several use cases. For example, a student enrolls in the university, and then immediately enrolls in three seminars [10].

- **The logic of methods.** Sequence diagrams can be used to explore the logic of a complex operation, function, or procedure. One way to think of sequence diagrams, particularly highly detailed diagrams, is as visual object code.

- **The logic of services.** A service is effectively a high-level method, often one that can be invoked by a wide variety of clients. This includes web-services as well as business transactions implemented by a variety of technologies such as CICS/COBOL or CORBA-compliant object request brokers (ORBs).

✓ **Sequence Diagram of Pay Bills**

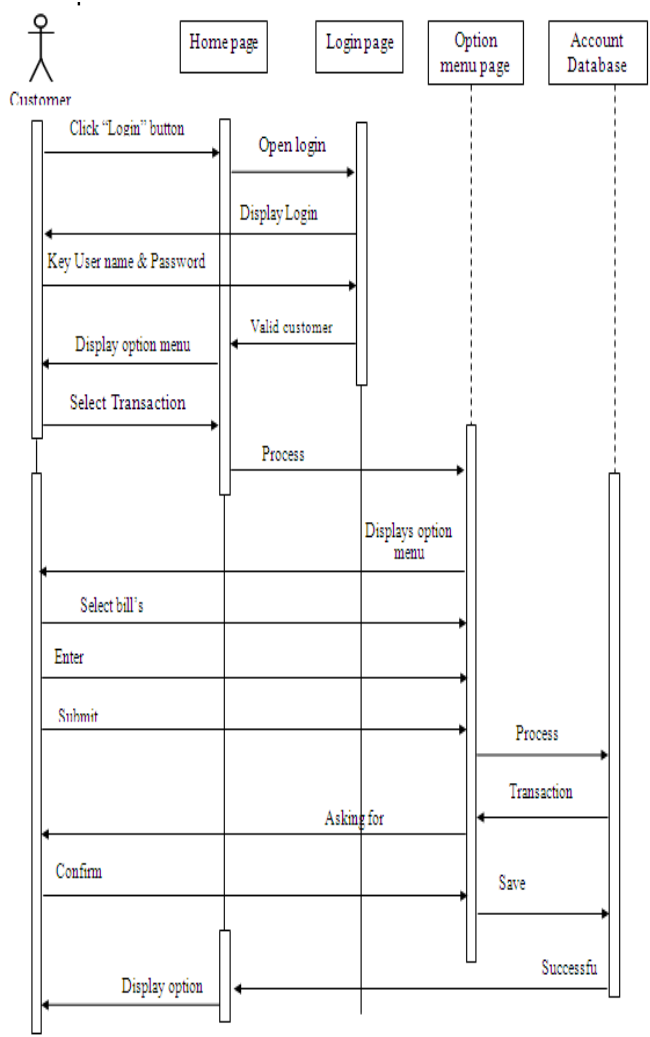

Fig 3. Sequence Diagram of "Pay Bills"

✓ **Sequence Diagram of Transfer Funds**

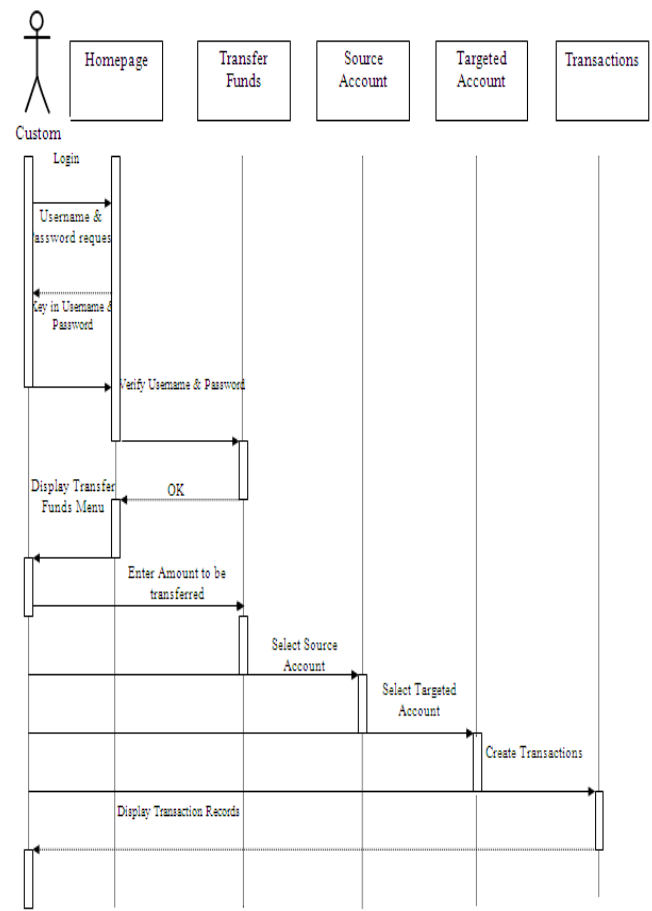

Fig 4. Sequence Diagram of "Transfer Funds"

✓ **Sequence Diagram of Cheque Services**

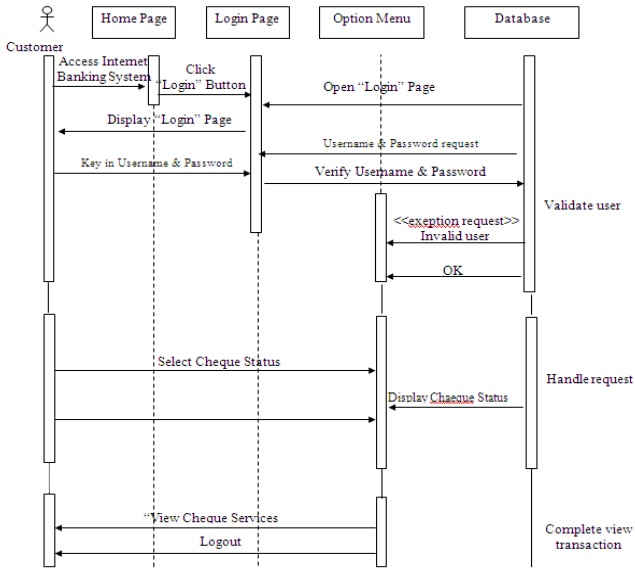

Fig 5. Sequence Diagram of "Cheque Services"

✓ **Sequence Diagram of View Account Transaction**

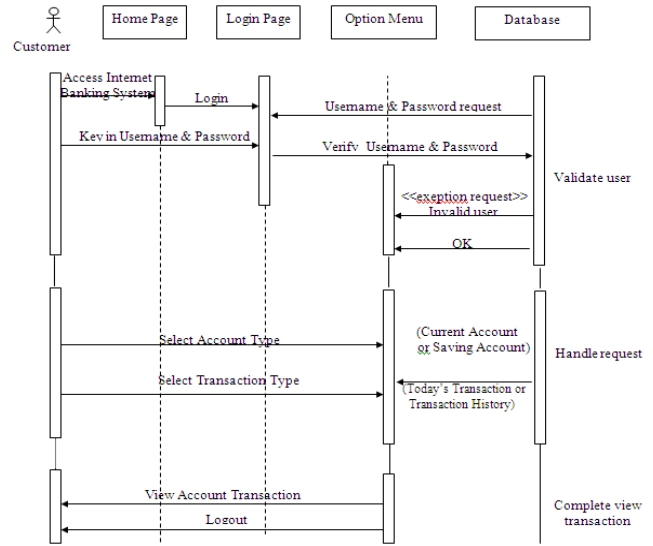

Fig 6. Sequence Diagram of "View Account Transaction" Use case

✓ **Sequence Diagram of Utilities**

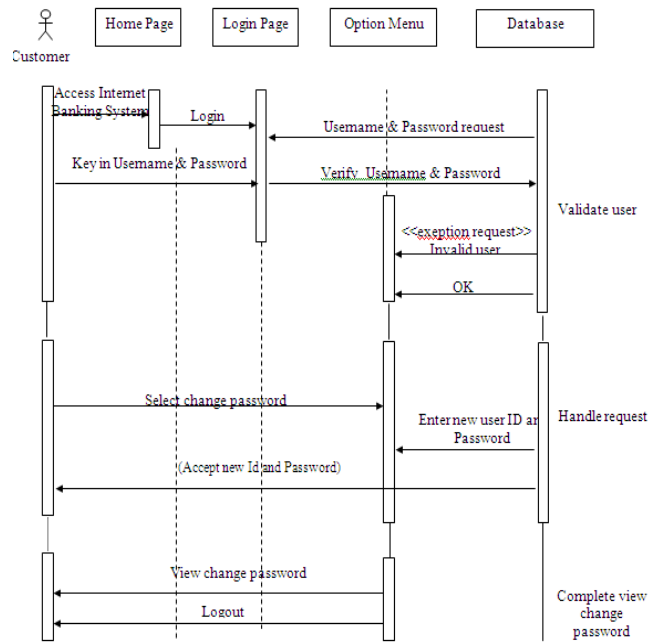

Fig 7. Sequence Diagram of "Utilities"

## 4. CLASS DIAGRAM

Static models of a system describe the structural relationships that hold between the Pieces of data manipulated by the system. They describe how data is parcelled out into Objects, how those objects are categorized, and what relationships can hold between them. They do not describe the behavior of the system, nor how the data in a system evolves over time. These aspects are described by various types of dynamic model. The most important kinds of static model are object diagrams and class diagrams. An object diagram provides a 'snapshot' of a system, showing the objects that actually exist at a given moment and the links between them. Many different object diagrams can be drawn for a system, each representing the state of the system at a given instant. An object diagram shows the data that is held by a system at a given moment. This data may be represented as individual objects, as attribute values stored inside these objects, or as link between objects [11].

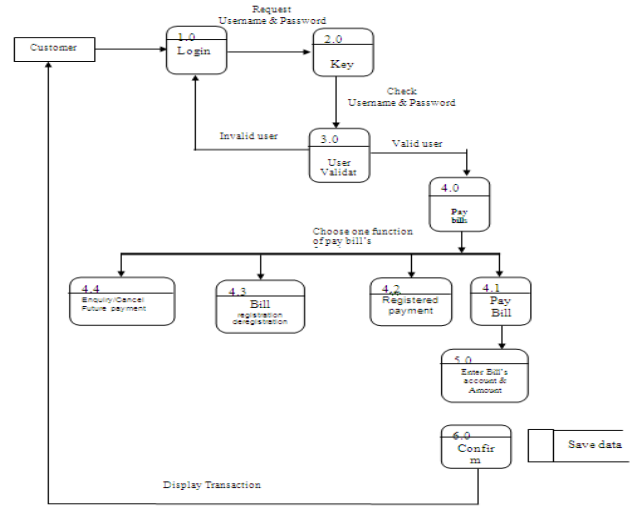

Fig 9: Data Flow Diagram of Pay Bills

✓ **Data Flow Diagram of Transfer Funds**

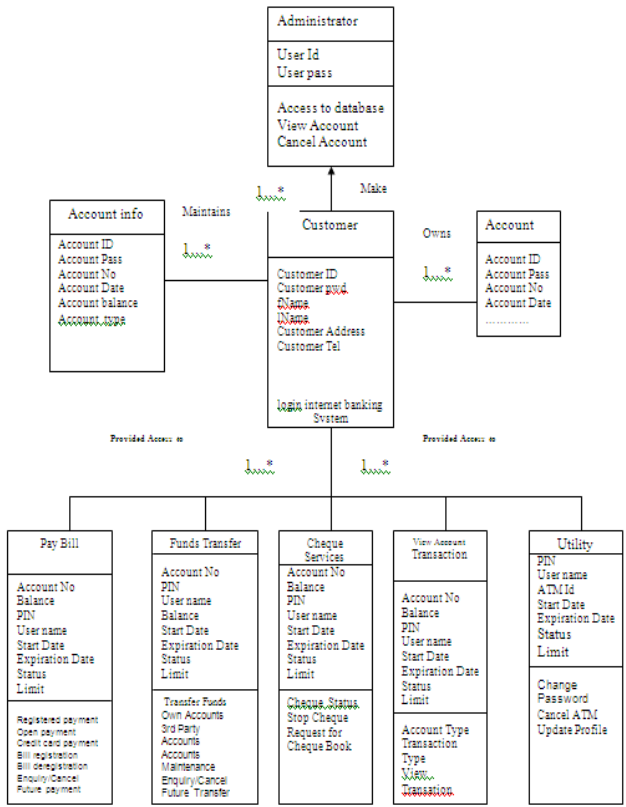

Fig 8. Class Diagram of Internet Banking System

## 5. DATA FLOW DIAGRAM

✓ **Data Flow Diagram of Pay Bills**

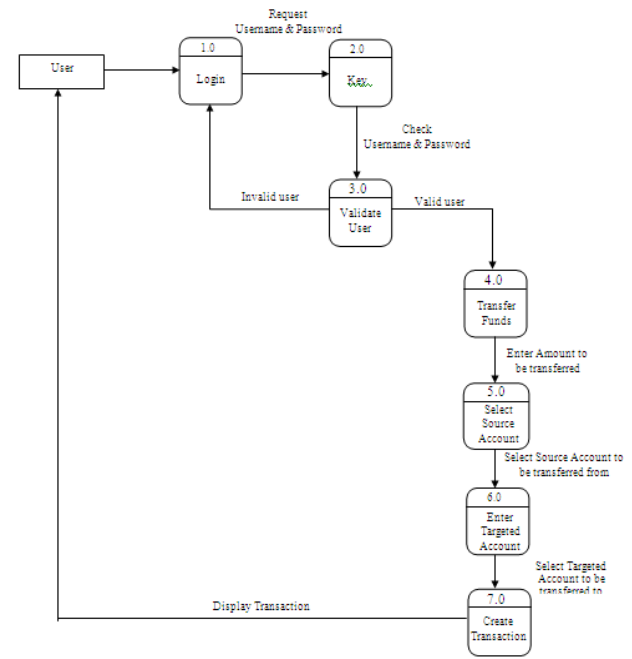

Fig 10. Data Flow Diagram of Transfer Funds

✓ **Data Flow Diagram of Cheque Services**

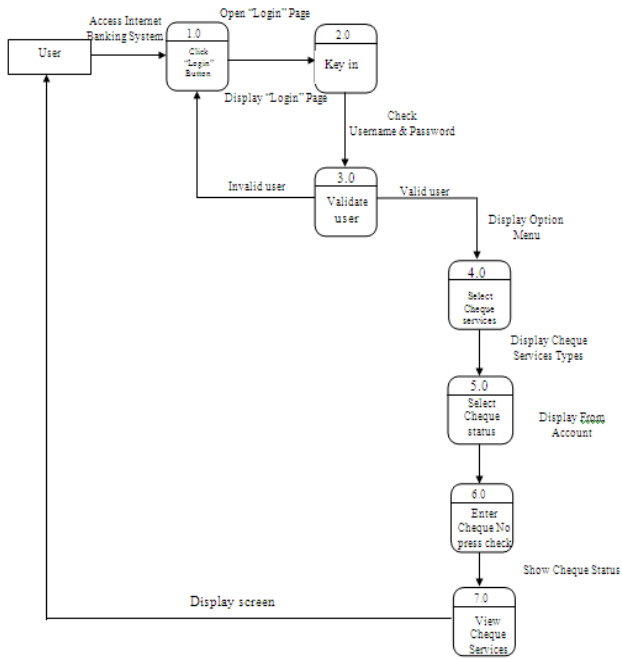

Fig 11. Data Flow Diagram of Cheque Services

- **Data Flow Diagram of View Account Transaction**

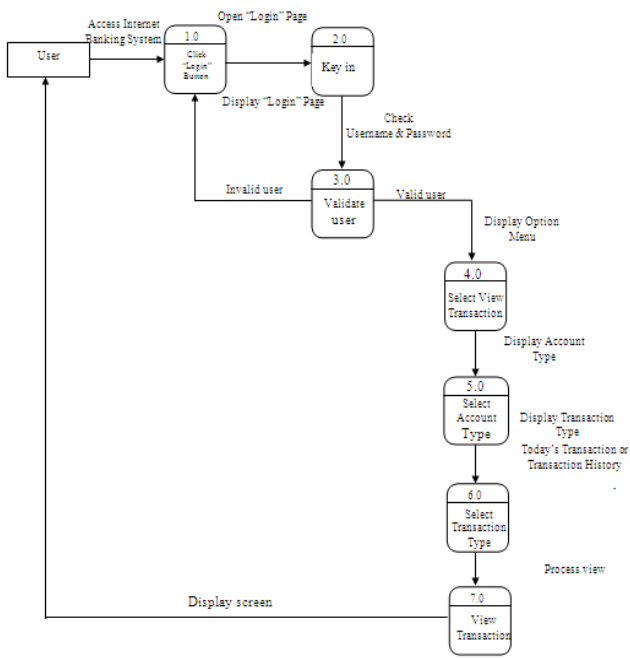

Fig 12. Data Flow Diagram of View Account Transaction

- **Data Flow Diagram of Utilities**

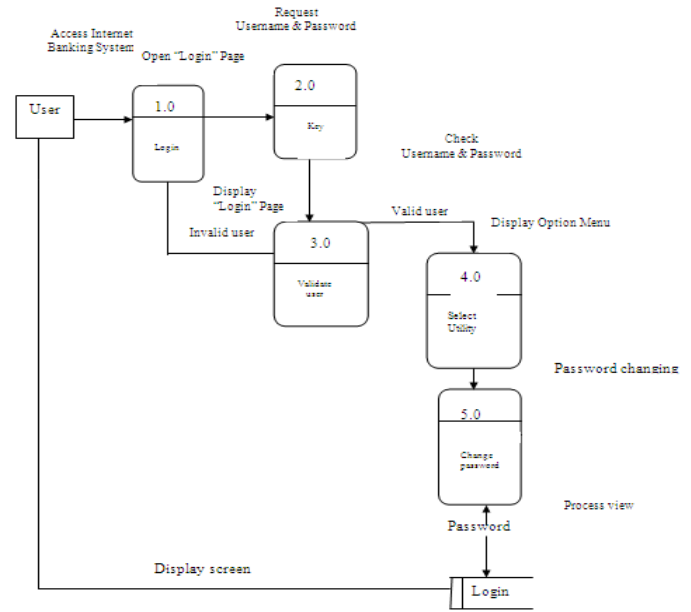

Fig 13. Data Flow Diagram of Utility

## 6. ARCHITECTURAL DESIGN

The architecture embodies the major static and dynamic aspects of a system. It is a view of the whole system highlighting the important characteristics and ignoring unnecessary details. In the context of our approach, architecture is primarily specified in terms of views of five models; the Use-Case model, Analysis Model, Design Model, Deployment model and Implementation model. These views show the "architecturally significant" elements of those models. The models have the following specific characteristics in our approach:  The Use-Case model shows the thematic use cases related to functionality associated with distribution The Analysis model illustrates how boundary, control and entity classes are associated with the thematic use cases identified in the Analysis. Remote Communication Control classes shown in this model are specializations of Control classes and represent the abstraction of components that deal with remote communication and distribution using CORBA (Common Object Request Broker Architecture).  The Design model shows the design classes that trace the specialized Remote Communication Control classes in analysis. Special attention is given to the interfaces provided by these design classes. We show how some of these are represented by IDL interfaces. The Implementation model describes how elements in the design model are implemented in term of components.  Finally, the Deployment model explains how CORBA-based components are assigned to nodes [12].

✓ **Two proposed Architectures for internet Banking System**

1. In a broader term, Architecture should not only focus on applications but it should also focus on Information Technology Service Management as a whole. It includes Technology, Processes, People and Information. Large data intensive system typically consists of a set of cooperating autonomous subsystems that brings the concept of multiple architectures within a system. In the following sections, we for large systems such as banking system to have an organizing concept that works for all modules within the system. In this way we can easily figure out the qualities that the proposed system should have in the required components. This will help to make the architecture more clear and understandable.

2. Within the architecture layer, we use different views to enhance the understandability of the architecture so that we can focus on particular concerns separately. We conceptual, logical and execution views, as described below.

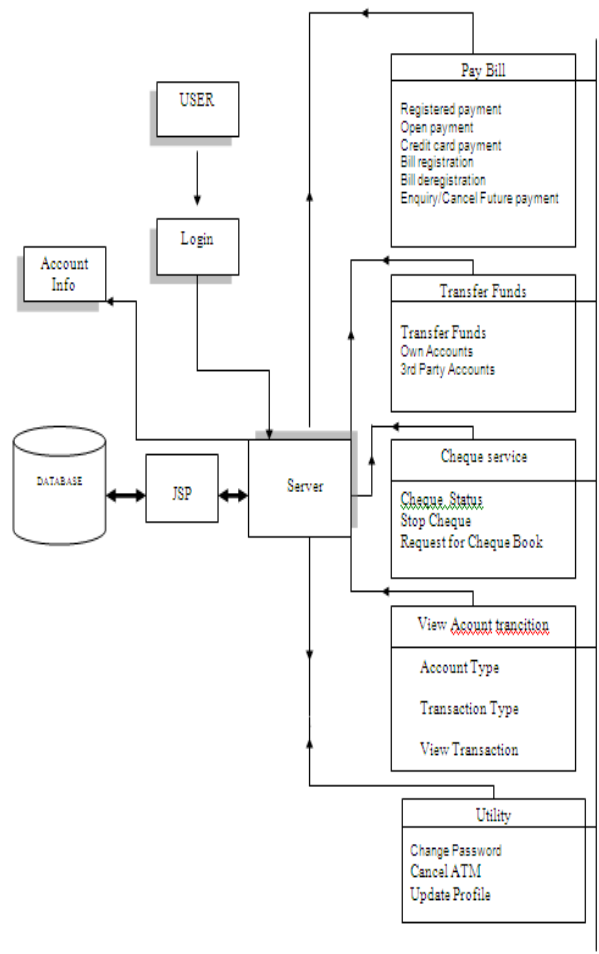

### 6.1 1ST ARCHITECTURE

Architecture Design involves in an early stage of the system design process represents the link between and processes. It involves identifying major system components and their communications The primary goals of architecture documentation are to Record the architects decisions in documentation. To meet this goal, the documentation must be complete and clear. Communicate the architecture. To meet this goal, consider what each stakeholder of the system needs to know, and what are the best procedures to convey what they need to know. The comprehensive architecture specification document can address this goal.

Architecture of internet Banking system:

The internet Banking Application is based on 3-tiered model. The Enterprise architecture For internet Banking Application is shown below.

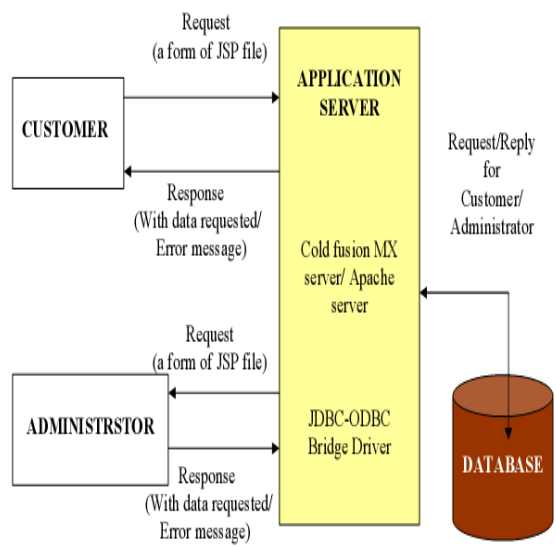

Fig 14. Architecture of Internet Banking Application

The 3-tiered architecture shown above has the following major components:

1. Client: There will be two clients for the application. One will be a web-based user-friendly client called bank customers. The other will be for administration purposes.
2. Application Server: It takes care of the server script, takes care of JDBC-ODBC (Open Database Connectivity) driver and checks for the ODBC connectivity for mapping to the database in Order to fulfill client and administrator's request.
3. Database: Database Servers will stores customer's and bank data

## 6.2 2ND ARCHITECTURE

The Conceptual Architecture identifies the high-level components of the system and the relationships among them. The purpose of this architecture is to direct attention at an appropriate decomposition of the system without discussing details. This view provides useful methods for communicating the architecture to non-technical audiences, such as management, marketing, and users. It consists of the Architecture Diagram (without interface detail) and an informal component specification for each component.

- ✓ **Modular Decomposition**

Modular decomposition is the process that the identified sub-systems are decomposed into sub modules.

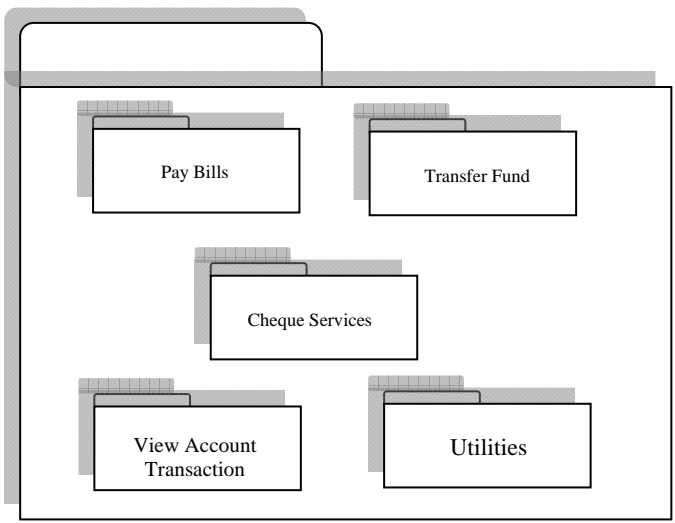

Figure 15: First Level of Decomposition

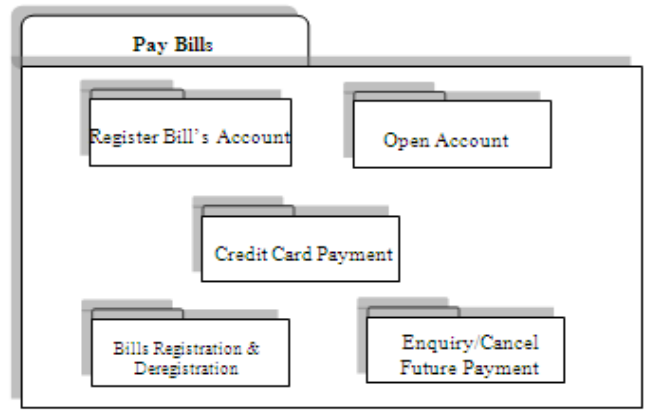

Fig 16. First Second of Decomposition

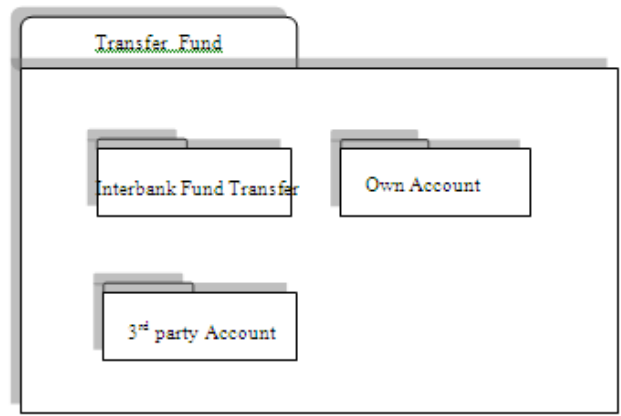

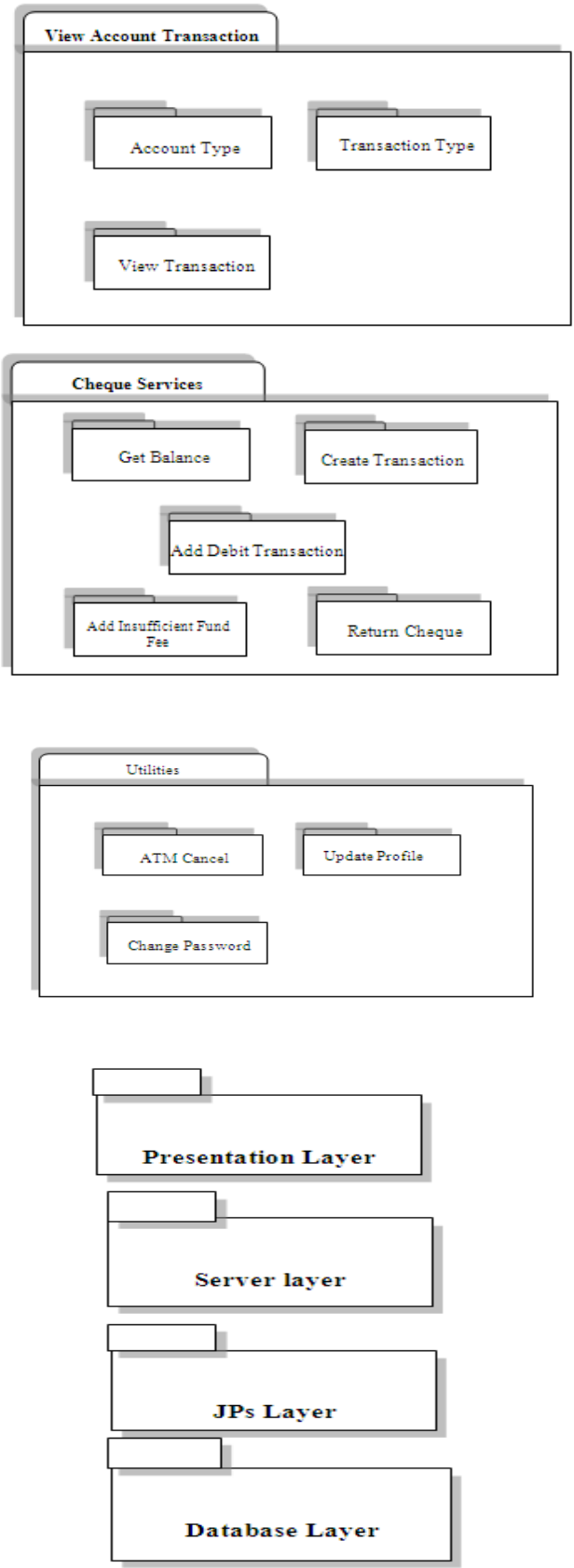

Fig 17. Layered Architectural

## 7. JUSTIFICATIONS OF THE PROPOSED ARCHITECTURES

The software architecture of a program or computing system is the structure of the system, which comprises software elements, the visible properties of those elements, and the relationships among them. Software architecture can be defined in many ways. In software architecture fundamental organization of a system embodies their relationships to each other and the environment in its components along with the principles governing design and evolution. An architecture can also be defined as a set of significant decisions about the organization of a software system, the selection of the structural elements, and their interfaces by which the system is composed, together with their behavior as specified in the collaborations among those elements. The composition of these structural and behavioral elements into progressively larger subsystems and the architectural style guides the organization, the mentioned elements, their interfaces, collaborations, and composition. Though Architecture drivers are not part of the architecture as such, the drivers that shape the architecture are important. They include: Architecture Vision: expressing the desired state that the architecture will bring about. Architectural Requirements: capturing stakeholder goals and architecturally significant behavioral (functional) requirements as well as system qualities (non-functional requirements) and constraints. Assumptions, Forces and Trends: documenting assertions about the current business, market and technical environment now and over the architecture planning horizon[13].

### 7.1 AVAILABILITY

Quality attributes of large software systems like banking systems are principally determined by the system's software architecture. In these systems the factors affecting the quality attributes such as performance, availability, and modifiability depends upon the overall systems architecture rather than on detailed system design, software development tools, algorithms, data structures and implementation. While describing Meta Architecture we will see cross domain aspects of infrastructure, semantics and service integration. In addition, we will also highlight the requirements of customers for internet banking system and how a layered approach reduces the complexity of design during the implementation of architecture for banking system[13],[14].

### 7.2 SECURITY

Security is possibly the most significant barrier to acceptance of IT services and digital services as a utility becoming absolutely crucial in a more dynamic often justify their creation by claiming that it supports and promotes certain qualities, often called non- functional qualities. These qualities including portability, reusability, performance, modifiability, and scalability are supposed to be automatically conferred on any system that is realized using the architecture. The achievement of non-functional qualities is attributable to many factors (such as coding styles, documentation, testing, etc.). However, larger the system, the more the achievement of non-functional qualities rests in a system's software architecture. In large systems a Meta Architecture approach will improve both the functional and non functional quality attributes [14].

## 8. CONCLUSION

Bank-Focused system model, though less risky, does not offer much when it comes to extending financial service outreach to the poor and unbanked. Both Bank-Led and Nonbank-Led system models offer a greater potential to achieve this objective. These system models, however, vary in their potential as well as risks. The decision as to which model must be adopted should be made after carefully weighing the risk-return tradeoff. A careful approach may be adopted to start with the less risky bank-led model and gradually adding more options as the players and stakeholders become more experienced. Once a model of branchless banking is decided upon, work towards creating an enabling regulatory environment for implementation of that model should start. Many components of such an environment are already in place if bank-led system model is adopted. However, Clear guidelines regarding various aspects of allowable activities should be issued to avoid uncertainties. Further, a forceful eradication of any unlawful and unauthorized services and offerings (generally provided by unlicensed players) - which     may sprout up - is a must to promote and safeguard the interest of genuine players and the overall system. Banking systems usually contains legacy systems along with very large database systems. For internet banking applications a large number of interfaces are incorporated to facilitate the customers especially in consumer banking applications. Handling of financial transactions requires taking care of various issues including authentication, consumer privacy, money laundering, liability for unauthorized transactions, security controls for safeguarding information and processing of third-party payments. The processes, services, channels, and available resources that are included in internet Banking can be categorized in different layers. The layered approach will help in defining different architecture to different sub systems and have meta- architecture above these architectures. This combination of architecture is capable of handling requirements of banking systems.

## ACKNOWLEDGMENT

This research was fully supported by "King Saud University", Riyadh, Saudi Arabia. The author would like to acknowledge all workers involved in this project that had given their support in many ways, aslo he would like to thank in advance Dr. Musaed AL-Jrrah, Dr. Abdullah Alsbail, Dr. Abdullah Alsbait. Dr.Khalid Alhazmi , Dr. Ali Abdullah Al-Afnan, Dr.Ibrahim Al-Dubaian and all the staff in king Saud University especially in Applied Medical Science In "Al-Majmah" for thier unlimited support, without thier notes and suggestions this research would not be appear.

**Hamdan Al-Anazi**: He has obtained his bachelor degree from "King Saud University", Riyadh, Saudi Arabia. He worked as a lecturer at Health College in the Ministry of Health in Saudi Arabia, and then he worked as a lecturer at King Saud University in the computer department. Currently he is Master candidate at faculty of Computer Science & Information Technology at University of Malaya in Kuala Lumpur, Malaysia. His research interest on Information Security, cryptography, steganography, Medical Applications, and digital watermarking, He has contributed to many papers some of them still under reviewer.

**Rami Alnaqeib-** he is master student in the Department of Information Technology / Faculty of Computer Science and Information Technology/University of Malaya / Kuala Lumpur/Malaysia, He has contribution for many papers at international conferences and journals

**Ali K.Hmood -** he is master student in the Department of Software Engineering / Faculty of Computer Science and Information Technology/University of Malaya /Kuala Lumpur/Malaysia, He has contribution for many papers at international conferences and journals

**Mussab alaa Zaidan -** he is master student in the Department of Information Technology / Faculty of Computer Science and Information Technology / University of Malaya/ Department /Kuala Lumpur/Malaysia, He has contribution for many papers at international conferences and journals.

**Yahya Al-Nabhani -** he is master student in the Department of Computer System and Technology / Faculty of Computer Science and Information Technology/University of Malaya /Kuala Lumpur/Malaysia, He has contribution for many papers at international conferences and journals.

.